\documentclass{kluwer}    
 
\newdisplay{guess}{Conjecture}

\def\lapproxeq{\lower.5ex\hbox{$\; \buildrel < \over \sim \;$}}
\def\simgt{\lower.5ex\hbox{$\; \buildrel > \over \sim \;$}}
\def\simpropto{\lower.2ex\hbox{$\; \buildrel \propto \over \sim \;$}}

\begin{document}                                                                                   
\begin{article}
\begin{opening}         
\title{Formation and Evolution of Disk Galaxies} 
\author{Joseph \surname{Silk}}  
\runningauthor{Joseph Silk}
\runningtitle{Formation of Disks}
\institute{University of Oxford\\Astrophysics,
Denys Wilkinson Building,\\
Keble Road,
Oxford OX1 3RH, UK}

\begin{abstract}
I review several of the current issues in the theory of disk galaxy formation. There is still much to be done, observationally and theoretically, before we can expect to approach an understanding of disk galaxies that is reliable enough to make robust predictions about the high redshift universe.
\end{abstract}
\end{opening}           

\section{Introduction} 

There is a well-accepted prescription for the formation of disk galaxies.
The dark matter context works well. Dark potential wells develop
hierarchically and acquire angular momentum via tidal torques with
neighbouring protohalos.  Baryons cool and dissipate their gravitational
potential energy, and conserve angular momentum to form a nearly
self-gravitating disk of size $\lambda R_h$, where $\lambda $ is the
dimensionless spin parameter acquired via nonlinear interactions and $ R_h$
is the halo virial radius. The disk cools and becomes gravitationally
unstable to massive cloud formation.  The clouds eventually themselves become
unstable and fragment into stars that form a disk with an exponential surface
brightness profile, a scale-length of a few kiloparsecs, and a 
central surface
density of order hundreds of solar masses per square parsec for Milky-Way
type spirals.  Star formation, fed and self-regulated by disk instability,
continues via ongoing accretion of gas into the disk from the halo reservoir
of gas and small satellites.

There are numerous observational probes of this simple picture.  The current
star formation rate in disks is measured via $H\alpha$ emission, effectively
probing the formation of massive stars. The disk gas fraction is measured via
HI and CO observations, and provides the fuel that drives disk star
formation.  Multicolour imaging provides a measure of the spectral energy
distribution, and probes the disk age over 10 Gyr.  Stellar absorption
features such as the Balmer spectral line index $H\beta$ measure ages over a
baseline of about 2 Gyr. Large samples of disk galaxies, most recently
utilizing the Sloan Digital Sky Survey, enable one to correlate surface
brightness with total stellar mass, galaxy radii and colours. 21 cm and
$H\alpha$ studies probe rotation curves, and allow one to explore the
Tully-Fisher relation. Galaxy scales, metallicities and correlations provide
fossilized glimpses of the galaxy formation epoch.  Observations at high
redshift can directly target the epoch of formation.  All of these probes
lead to strong constraints on the basic disk formation model.

Theory provides a well-accepted  framework
for the hierarchical formation of dark matter halos. However it is far weaker when it comes to the formation
and evolution of the star-forming components of galaxies. In this review, I will
concentrate on disk galaxies, and discuss the problems that have arisen with 
some of the proposed solutions. Much of our understanding centres on the concept of self-regulation of the global star formation rate. Unfortunately, some of the key ingredients are poorly known.

\section{Disk formation}

The dark matter that dominates the matter budget of the universe, and in particular the galactic mass budget, is considered to be non-dissipative.
While the precise nature of the dark matter remains elusive, the hierarchical formation of non-dissipative  dark potential wells in which baryons dissipate and condense into stars,  has provided a successful model  for many elements of the large-scale structure of the galaxy distribution \cite{White}.
Galaxy clustering can be explained in such a framework, as can the properties of the dark halos that are inferred, for example, from rotation curves, and 
of the intergalactic medium where neutral baryons in the form
 of the Lyman alpha forest provide a powerful probe of the weakly nonlinear regime. Tidal torques are generated by nonlinear interactions between neighbouring fluctuations and nascent halos. If the baryons conserve angular momentum as they dissipate kinetic energy  and contract, disks 
spanning  the observed
size  distribution  are formed. The distribution of angular momentum is not such a good match, as the theory predicts far more low angular momentum gas than is seen. Once the baryon disk is self-gravitating,  it is gravitationally unstable and fragments into stars. This general overview of disk formation can match many of the observed properties of disks
provided that disk star formation is an inefficient process.  Observed disks are still gas-rich and star formation extends over 50 or more disk dynamical time-scales.

\section{Global star  formation in disks}

Disk star formation can be understood via a hybrid model
that is a combination of phenomenology and gravitational instability theory.
The low efficiency  of star formation is determined by the gravitational instability
of a cold gas-rich disk, resulting
in the formation of warm molecular gas clouds which in turn fragment into stars.
Cloud collapse and star formation  are enhanced by cloud coalescence and growth, as the clouds orbit the galaxy.
This process is modulated and amplified by the spiral density wave pattern of the differentially rotating disk. The instability in a gas disk is quenched by the Toomre criterion, when the Toomre parameter, $Q\equiv \mu_{crit}/\mu_{gas}$,
becomes larger than unity. Here the critical surface density is, to within a numerical factor of order unity, given by $\kappa\sigma_{gas}/G$ with   $\kappa$ equal to the epicyclic frequency and $\sigma_{gas}$ being
the gas velocity dispersion. For a disk consisting of stars and gas, the criterion must be slightly modified, but it is essentially
the coldest component, the gas, that drives the instability,
provided that the gas surface density
$\mu_{gas}\simgt\mu_\ast\sigma_g/\sigma_\ast,$
where $\mu_\ast$ is the disk stellar  surface density and $\sigma_\ast$
is the stellar velocity dispersion.

This scheme is used to derive an empirical star formation rate, the
Schmidt-Kennicutt law, which has been applied to fit a large sample of
star-forming disk galaxies. Beyond the radius where the
azimuthally-averaged surface density drops below the critical value,
the disk is stable to cloud formation. The inferred star formation
rate, if one assumes that the clouds once formed are unstable to star
formation within a disk rotation time or less, can be approximated by 
$$\dot\mu_\ast=\epsilon\mu_{gas}^n\Omega(Q^{-2}-1), $$ for $Q<1,$ with
$n\approx 1.$ The instability requires the disk to be in differential
rotation, at rate $\Omega(r)$.

The predictions of such a semi-empirical model are straightforward:
star formation occurs inside-out in disks, quenching below the
threshold surface density of cold gas ($HI$ or $H_2$) of a few solar
masses per square parsec, and star formation decays monotonically as
the gas supply is exhausted. The duration of star formation can be
extended if gas infall occurs from the halo.

Broadly viewed, such a model works reasonably well from the point of
view of chemical evolution (cf. Prantzos, these proceedings).
However detailed recent studies of disk star formation suggest that there are significant omissions in the underlying physics.
For example, there is  the realization that the radial dependence of
disk star formation  is not always well modeled  by the empirical law
\cite{Ferguson}. In some galaxies,
star formation occurs below the threshold surface density. This is most likely due to the non-axially symmetric distribution of the gas. Locally the gas surface density may be high, as for example in spiral arms.

Another hint of the need for a more complex theory comes from studies of the age distribution of disk stars.
In the Milky Way, the star formation rate history is seen to be non-monotonic
 \cite{Rocha}.
In fact, studies of chromospheric age indicators reveal a series of modest star bursts. A similar pattern is found in other nearby galaxies where stars can be resolved, and the Hertzsprung-Russell diagram can be used to study the star formation history.
Presumably, infall is occurring in a non-uniform way, for example via mergers of satellite galaxies. Studies of the metallicity distribution of old disk stars in the solar neighbourhood require early infall in order to account for the paucity of old metal-poor stars.

The underlying logic of dissipative disk formation in weakly interacting
halos of cold dark matter seems compelling.  Star formation is
inefficient in disks because of self-regulation.  The disk forms via collapse
in the dark halo of gas that has acquired some angular momentum via tidal
torques between neighbouring halos. The dimensionless angular momentum
parameter is initially $v_{rot}/\sigma \approx 0.15,$ and contraction by a
factor of order 10 in an isothermal dark halo of virial radius 100 kpc that
dominates the gravity results in a disk of scale around 5 kpc if specific
angular momentum is approximately conserved.

The gas disk is cold and gravitationally unstable, forming giant cloud
 complexes that aggregate gas and are in turn unstable to
 fragmentation. Feedback both via dynamical heating of the stars and by
 supernova remnant interactions heat and stabilize the system against further
 star formation, until further gas infall drives further gas cooling.

Because of the bottom-up nature of the clustering hierarchy, driven by the
approximately scale-invariant primordial density fluctuation spectrum, the
density fluctuation amplitude scales with mass as $\delta\rho/\rho \propto
(1+z)^{-1}M^{-(n+3)/6}$ with $n\approx -1.5$ on galaxy scales. This means
that massive halos form after lower mass halos, with the virialization
redshift scaling as approximately $M^{-1/4}$. There is as much mass in the
low mass halos as in the massive halos, with the number of halos scaling as
$dN/dM\simpropto M^{-2}$.

There are serious deficiencies in the model that remain when disk
formation is incorporated by such simple rules into semi-analytic galaxy
formation theory that builds on numerical simulations of dark matter
clusters.

\section{The dark side of disk formation}

Several fundamental problems have arisen in detailed attempts to implement
disk formation. The baryons undergo excessive cooling.  The highest
resolution simulations to date have revealed that as a consequence of the
unresolved clumpiness of the dark matter, most of the baryonic angular
momentum is lost to the dark halo by dynamical friction of infalling baryon
clumps.  The final specific angular momentum is only twenty percent or less
of what is observed. The angular momentum distribution also does not match
that of dark halos if the baryon angular momentum distribution tracks that of
the dark matter, there being far too much low angular momentum gas at small
galactocentric radii in the models.  Even if angular momentum is assumed to
be conserved by the baryons, the final disk in a Milky Way-like galaxy is
about twice as massive as observed.

One issue is generic to hierarchical formation models with a specified efficiency
of turning gas into stars. The most massive systems form late, and hence are
expected to be younger and therefore bluer than less massive disks. The
opposite trend is seen in the disk colour-magnitude relation
\cite{vandenBosch}. 

A further difficulty arises from the substructure in the dark halos that
results in the formation of many satellite halos.  These are not observed as
stellar systems, for example in the distribution of Local Group dwarf
galaxies.
Another problem concerns the density distribution of the dark matter halos.
The dark matter halos seem to be too concentrated compared to actual
galaxies.  Cusps are predicted in the halo cores that are not seen in nature,
and the ratio of dark mass to baryonic mass is excessive in the vicinity of
the baryon-dominated disk.

 Several galaxies have been modelled in detail and
illustrate these various trends. The Milky Way has the best studied rotation
curve.  Binney and Evans combined the inner rotation curve with microlensing
data to infer that the observed microlensing optical depth towards the
galactic centre requires so many stars that one has difficulty in
acommodating a diffuse cold dark matter component. No more than 10 percent of
the dark matter within 5 kpc can be diffuse and non-baryonic
\cite{Binney}. In particular, the NFW profile ($\rho\propto r^{-1}$ for
$r<r_s$ with $r_s\sim 0.1r_{200}$) is not allowed.  Indeed, the highest
resolution simulations require an even steeper dark matter cusp; $\rho\propto
r^{-1.5}$.  This certainly would not allow enough bulge and inner disk stars
to give a reasonable microlensing optical depth.

A complementary study by Klypin et al focuses
on the outer Milky Way rotation curve. They argue that up to half of the mass within 10 kpc may be contributed by the dark halo, thereby allowing a NFW profile \cite{Klypin}.
However this conclusion comes at a cost. Only 60 percent of the baryons inferred by the model to be within the virial radius can be accomodated in the
bulge, stellar halo and disk, including interstellar matter.
Most of the observed mass of course is in stars, which are measured directly
for stars of order a solar mass or larger, and via 
rotation curve modelling, diffuse infrared emission
and bulge microlensing for the least massive stars.
All disk formation  and evolution simulations seem
to run into a similar difficulty,
as for example  in a  recent discussion of colour evolution \cite{Westera}.

Several of these issues are manifestations of the overcooling problem that is generic to
hierarchical structure formation models. The cooled baryon fraction
in cluster simulations is about 30 percent, whereas the global CDM prediction is about 5 percent for the baryon fraction, which for $\Omega_m\approx 0.3$
predicts 15 percent for rich clusters, as indeed is observed. About
half  of the cooled baryons remain on the peripheries of clusters
as the warm/hot intergalactic medium at $T\sim 10^5-10^6$ K.
Numerical SPH  simulations find such a WIGM,
although the physics of how this gas is heated yet remains outside the cluster virial radius is not completely clear. Shock heating is obviously important, but there are numerical issues that  need to be clarified. The WIGM however only acccounts for about 30 percent of the total baryon content of the low redshift universe.
The cluster x-ray luminosity-temperature correlation, which monitors the mass in virialised gas,
does not satisfy the simple scaling expected in hierarchical collapse models, but rather suggests that preheating, preferentially in the lower mass clusters, may have helped to steepen the luminosity-temperature correlation from
$L_x\simpropto T^2$ to $L_x\simpropto T^2$.

Another probe  of dark matter in reasonably luminous disks comes from
 studying the kinematics of barred galaxies. Debattista and Sellwood argued
 that a maximal disk is required, eg for NGC 3198,
 in order to maintain a self-gravitating bar
\cite{Debattista}. On the other hand, Kranz et al. find  varying results for the disk mass fraction by modelling the detailed kinematic structure, for example the (unbarred) 
galaxy NGC 4254 requires a submaximal disk \cite{Kranz}.

The colour-magnitude relation (B-K versus K) reveals 
another difficulty \cite{vandenBosch}. The more luminous galaxies, the redder they are,
indicative of less vigorous recent star formation. SEmi-analytical
disk models, even when feedback is included, give the inverse correlation. This is another manifestation of the overcooling problem.
The oldest early-type galaxies are found to be the most massive, and these have the highest $[\alpha/Fe]$,
indicative of the shortest star formation time-scales.
The opposite trend is expected in hierarchical models \cite{Thomas}.

\section {Proposed solutions}

The baryon fraction in stars may be understood if approximately half the
baryons that initially cooled within the virial radius have been expelled.
Supernova feedback seems incapable of driving such a wind in the early
gas-rich phase of galaxy formation because of the significance of cooling.
However there are indications that vigorous winds are seen in the Lyman break
galaxies, as inferred from the inverse P Cygni line profiles in template
spectra and from the Lyman alpha forest suppression found within a Mpc or so
of the Lyman break galaxies.  In nearby galaxies where so-called superwinds
are found, the observed mass outflow rate is comparable to the star formation
rate, e.g. in NGC 253 \cite{Strickland}.  Observations suggest that strong winds can be driven from typical
galaxies, yet numerical simulations find that winds are stalled for the more
massive halos.

What physics has hitherto been omitted from the simulations?  The simulations
are gravity-driven. Star formation may be, in extreme situations,
pressure-driven. This could result in strong positive feedback via for
example shock-triggered  massive star formation and generate 
enhanced rates of  correlated  supernovae.
that would in turn drive a transient wind. 

Starbursts are known to drive winds.
One mechanism appeals to bar formation.
A merger creates a transient
bar. The spin-down and dissolution of this bar torques and accelerates the
gas which responds by an enhanced rate of dissipation and angular momentum
loss, for example via cloud-cloud collisions. 
Bars are known to drive luminous starbursts, and minor mergers may drive
ultraluminous starbursts, the ultimate sources of superwinds.

The gas-rich environment provides a plausible site for 
supermassive black hole
formation. This  phenomenon 
 appears to occur contemporaneously with spheroid formation, as
inferred from the remarkable correlation between central black hole mass and
spheroid velocity dispersion. SMBH-driven outflows, manifest as a
SMBH-energized component of the overall luminosity, may be important in early
phases of galaxy formation. At least one explanation of the observed
correlation appeals to outflows as the self-regulatory mechanism.

While the formation of structure in halos building up from clouds of weakly
interacting dark matter is reasonably well understood, the formation of the
luminous components of galaxies is founded on simplistic assumptions.  A more
realistic treatment of star formation in forming galaxies must ultimately
modify many of the predictions that are failing to adequately confront the
observations.

Feedback from supernovae is one solution  that is being widely discussed.
Supernova remnants deposit substantial energy into the interstellar medium,
and are responsible for the observed multiphase structure of the interstellar
gas. Some nearby low mass star-forming galaxies show a porous structure in
the interstellar gas, where many overlapping shells and bubbles are seen that
demonstrate that supernova input is playing an important role in pressurizing
the interstellar gas. Some display an anti-correlation between X-ray and
$H\alpha$ emission, indicative of supernova-driven bubbles.
The microphysics of the  interface
between the supernova-heated gas and the cold
interstellar medium is inadequately resolved by the simulations. Plausibly,
there are surprises in store.

Feedback reduces the efficiency of star formation.  This is important for
understanding the longevity of disk star formation.  However attempts to use
feedback in the phase of disk formation have had mixed success.  A basic
problem is that feedback delays star formation: if this works too well, all
stars are young, in contrast to what is found in the outer parts of nearby
disks such as M31.  Numerical simulations that incorporate feedback find that
disks are still too small, although possibly by only a factor of 2
\cite{sommer}.

If efficiency of star formation is to account for the observed
colour-magnitude relation, then one needs a systematic reduction in efficiency
of star formation with decreasing galaxy mass. Some evidence for this is found in the correlations between stellar surface brightness and total stellar mass in the SDSS study of 80,000 early and late-type galaxies \cite{kauffmann}.
In ellipticals,
such an effect would help to simultaneously account for the trend of increasing $[\alpha/Fe]$ with galaxy mass, if the most massive galaxies are the oldest and formed stars most rapidly. However no detailed implementation has been made of the relevant physics, in large part  because one needs to incorporate a multiphase interstellar medium.

\section{Analytic disk galaxy formation}

The brute force computational approach will not resolve the outstanding problems in disk galaxy formation. What seems to be needed are further analytical insights that will allow refinement of the simple prescriptions for star formation. One such approach has come from studying turbulence-driven viscous evolution of differentially rotating disks. Recent investigations of star formation suggest that turbulence plays an important role in accounting for the longevity of star-forming clouds and their fragmentation into stars. The Jeans mass in a typical interstellar cloud greatly exceeds the stellar mass range.
It is likely that the gravitational instability of galaxy disks is a primary source of interstellar cloud turbulent motions, supplemented on small scales by supernova feedback. Of course, these drivers of turbulence are coupled together, since the rates of star formation and star deaths are controlled by global gravitational instability.
In effect, differential rotation is the ultimate source of the turbulence.

  A promising hypothesis is that turbulent viscosity, by regulating
the gas flow, controls the star formation rate, and indeed that the
star formation time-scale is given by the time-scale for the viscous
transfer of angular momentum
\cite{SilkN}. On the scale of molecular clouds, such an ansatz is
reasonable, since one has to shed angular momentum in order to form
stars. Magnetic fields are the common culprit in conventional
star-forming clouds, but in protogalactic disks one most likely has to
appeal to another source of angular momentum transfer. Turbulent
viscosity is capable of fulfilling this role. Indeed, the resulting
disk has been shown to generically develop an exponential density
profile
\cite{Lin}. In infall models, the initial angular momentum profile
determines the final disk scale length if angular momentum is
conserved. However as found in high resolution simulations,
some 90 percent of the baryonic specific angular momentum is lost
to the dark halo, and there is no preferred solution for disk sizes.
In viscous disk models,
the scale length is set by the competition between viscosity-driven
star formation, that freezes the scale length once stars form, and
dynamical friction on the dark matter, that competes for the same
angular momentum supply.
The characteristic viscous scale is  determined by the cloud mean free
path between collisions, itself comparable to the disk instability
scale that drives the turbulence, and
in combination with the residual rotation rate, provides the ultimate
constraint on disk scales.

Another byproduct of the viscous disk model is the gas fraction
\cite{Silk}. The viscous redistribution time-scale, and hence the star
formation time-scale, is
$$t_\nu\approx t_{sf}\approx r_d^2/\sigma_{gas}\ell,$$ where $\ell$ is the
cloud mean free path (of order the disk scale height), and $\sigma_{gas}$ is
the cloud velocity dispersion. Disk instability yields $\sigma_{gas}\approx
\Omega\ell.$ The star formation efficiency, if determined by supernova
feedback and approximate conservation of momentum, is $\epsilon\approx
\sigma_{gas}/v_{SN},$ where $v_{SN}$ is the specific supernova momentum per
unit mass injected into the interstellar medium. Here $v_{SN}\equiv
E_{SN}/m_{SN}v_c,$ where $m_{SN}$ is the mass in stars per supernova and
$v_c$ is the velocity of transition of a remnant to approximate momentum
conservation.  The characteristic star formation time may then defined to be
$$t_{sf}\approx \frac{M_\ast}{\dot M_\ast} \approx\frac{M_\ast}{\dot
M_{gas}}\epsilon^{-1}\Omega^{-1},$$ where $\Omega(r)$ is the disk rotation
rate at radius $r$, $M_\ast(r)$ is the instantaneous stellar mass and
$M_{gas}$ is the gas mass. A steady state is reached in which the gas
fraction is
$$\frac{M_\ast}{ M_{gas}}\approx \frac{\sigma_{gas}v_{SN}}{v_{rot}^2},$$ and
the disk scale length is
$$r_d=\ell \sqrt{\Omega t_{sf}}.$$ Thus the disc scale depends both on
cosmology and on local conditions.  

The inferred present epoch numbers are
plausible: 
for $M_\ast \approx 6\times 10^{10} \rm
M_\odot, $ one finds $\dot M_\ast \approx \rm 3M_\odot/yr$
and $M_{gas}/M_\ast \approx 0.1.$ 
 Also, one has now $\sigma_{gas}
\approx 10 \rm km s^{-1},$ $v_{SN} \approx 1000 \rm km s^{-1},$ and
$\ell\approx 0.3$kpc. At disk formation, 
one expects that $M_{gas}/M_\ast \approx 1$,
$t_{sf}\approx 10^{9}$ yr, and  $\ell \approx 1$ kpc, appropriate to the
protodisk. These values result in
a  stellar disk scale-length  $r_d\approx 3$
kpc.
It is
encouraging that simple analytic estimates come out with reasonable numbers
for gas and stellar disk scales and gas fraction. Whether such a simple model
survives incorporation into  3-dimensional simulations of disk formation in
the presence of a live dark matter halo and energetic winds remains of course to be seen.

\section{Spheroid formation}

Spheroids are an integral component of practically all disk galaxies. Studies of spheroid formation are complicated by the fact that there is no global theory of star formation even at the semi-empirical level that works moderately well for cold gas disks.

Small spheroids most likely form by secular evolution of bar-unstable disks. However it is not possible to form massive spheroids by secular evolution.
Galaxy mergers are undoubtedly the major trigger for formation both 
of massive spheroids in early-type disk galaxies and of elliptical galaxies.
It has been realized from the earliest simulations of mergers that gas-rich precursors 
are required  in order to attain the high central densities of massive spheroids. The gas dissipates, forms stars that in turn self-enrich more gas, ending up with the metal-rich nuclei characteristic of massive ellipticals. This must have happened with high efficiency, since spheroidal stellar populations are characteristically old.
The necessarily high star formation rate constitutes a star burst, defined simply by the requirement that the mean star formation rate is inferred to have been much higher, perhaps by two orders of magnitude, than the mass of stars divided by a Hubble time.

Nearby starbursts are well studied. Typically, starbursts are rare in the
nearby universe but increasingly common at high redshift. Star formation
rates of up to 1000 M$_\odot$ yr$^{-1}$ are inferred.  The most extreme star
formation rates are invariably associated with evidence for an ongoing
merger. Most of the star formation generated by gas concentration triggered
in a merger is shrouded by dust, and most of the radiation is absorbed and
reemitted in the far infrared. Simulations demonstrate that the mergers of
gas-rich galaxies efficiently drive the gas into the central kiloparsec of
the merged galaxy by a powerful combination of tidal torquing on the gas due
to the merged transient stellar bar that results from the merger, which
removes angular momentum and compresses the gas, followed by strong
dissipation of energy by radiative cooling of the gas.

Ultraluminous infrared starbursts are likely to be the sites of ongoing
spheroid formation. This conjecture is supported by near-infrared
observations of post-starburst galaxies, where the characteristic de
Vaucouleurs profiles found in spheroids can be recognized in the newly formed
stars.

There is an intriguing complication. Spheroids have been found
 to be
intimately linked with supermassive black holes.  The tight correlation 
observed
between black hole mass and spheroid velocity dispersion covers the range of
black hole masses $10^6$ to $10^9$ M$_\odot$.  and encompasses stellar
spheroids as small as that of the Milky Way ($\sim 10^9$ M$_\odot$) to those
as massive as M87 ($\sim 10^{12}$ M$_\odot$).  The clear implication is that
the formation of spheroids and supermassive black holes was
contemporaneous. If the spheroid formed by a merger, the strong central gas
enhancement provides an ideal environment for supermassive black hole
formation. The enormous amount of binding energy released 
as the SMBH formed would impact the
protogalactic gaseous environment, as would the 
ensuing effects of energy released by
infall into the newly formed black hole.  Outflows would be a natural
outcome, and these could help stimulate further star formation.

Observationally, it is unclear what role quasar-like activity plays in the
energetics of the far-infrared and submillimeter galaxies that are undergoing
luminous starbursts. In the nearby case of Arp 220, it is apparent that star
formation and buried quasars play comparable roles in accounting for the
observed luminosity.

\section{Conclusions}

There seem to be two distinct modes of global star formation. An inefficient,
quiescent mode that is self-regulated by feedback occurs in galactic disks.
A violent, efficient starburst mode is triggered and fueled by merger-induced
tidal torques, and accounts for the formation of the massive stellar
spheroids.  The disk mode of star formation most likely also accounts for the
low mass spheroids, which are generated dynamically via secular evolution of
disks.

Semi-analytic galaxy formation modelling has not yet incorporated the full
richness of star formation phenomenology.  There are five critical issues
currently confronting semi-analytic theory, and improvements in star
formation modelling and the dynamical coupling of baryonic and non-baryonic
matter will be necessary to address most of them.
\begin{itemize}
\item Perhaps the most innocuous of the problems is the overproduction of
substructure, and in particular the predicted abundance of dwarf galaxies.
This is likely to be resolved at least in part by early photo-ionization at
redshifts above 6, before the reionization of the intergalactic medium was
complete. Early photo-ionization of dwarfs with velocity dispersion less than
about 30 km/s suffices to eject most of the gas before the bulk of the star
formation has occurred. Because this happens early, there is a proximity
effect: only the most widely separated dwarfs, which experience a lower UV
flux, retain their gas and survive as stellar systems. One can both account
for the paucity of observed dwarfs and their mean separations from the parent
galaxies. What is not yet clear is whether the number of SMC-like systems
predicted is consistant with observations, nor more seriously perhaps, is
whether reconciling the low dwarf abundance in the Local Group allows one to
also understand the apparent surfeit of dwarfs observed in some nearby galaxy clusters. 

\item The baryon overcooling problem is more serious. The factor of about 2
excess predicted for disk baryons can be resolved either by ejection or by
hiding the gas.  Winds can occur if outflows from massive galaxies are
stronger than predicted by simple disk modelling and simulations.  The Lyman
break galaxies possibly indicate the effectiveness of early winds, but it is
likely that these galaxies end up as E or S0 galaxies, as inferred from their
spatial clustering. Substantial amounts of gas could be hidden in the form of
dense cold clumps of $H_2$, or even in diffuse $H_2$ in the outer disk,
where $H_2$ may be more readily hidden.
\item Disk scale lengths can possibly be explained via turbulent
feedback. Winds may play a role here too, if the predominantly low angular
momentum gas, which ends up in the central core, is preferentially
expelled. A supermassive black hole-driven outflow may provide a possible
driver for the nuclear wind.
\item Disk colours, and indeed spheroid colours, present another difficulty,
as exemplified by the colour-magnitude relation.  Disks, whose global colours
are characteristic of intermediate age populations, have stellar populations
that are systematically too red, and too old, at large masses.  For
ellipticals, the problem is somewhat different. The colour-magnitude relation
is driven primarily by metallicity, with there being little evidence for any
significant age spread in cluster galaxies \cite{Vazdekis}. The stellar
populations are predominantly old. Both massive and low mass ellipticals have
similar ages.  Either the low mass systems formed stars earlier but less
efficiently, hence reducing their effective age, or gas accretion occurred to
produce a similar effect.  The alpha element ratio enhancement for the cores
of massive ellipticals supports an early and efficient duration of star
formation for these systems.  Field ellipticals and S0s however show a
substantial age spread.
\item The issue of a cusp in the dark matter profile represents a contentious
issue around which the observations have not yet converged.  It may be that
dynamical effects, such as black hole mergers, provide the best means of
scouring out the cusp.  The related matter of
dark matter concentration seems to vary between galaxies, 
and  there are hints that it varies between barred and unbarred galaxies.
This suggests again that a complex dynamical history may be partly to blame.
\end{itemize}
\section{Acknowledgements}

I am indebted to my group at Oxford for many discussions of galaxy formation,
and in particular for conversations pertinent to this review  with
Julien Devriendt, Ignacio Ferreras, Adrianne Slyz and
James Taylor. Part of this review was completed at the Institut
d'Astrophysique de Paris, where I am grateful to the Director,
 Bernard Fort,
for his kind hospitality.
I also thank N. Prantzos for useful discussions.

\end{article}

\begin{thebibliography}{}

\bibitem[\protect\citeauthoryear{Binney \& Evans}{2001}]{Binney}
Binney, J. and Evans, N.  2001, MNRAS, 327, L27.
\bibitem[\protect\citeauthoryear{Debattista \& Sellwood}{2000}]{Debattista}
Debattista, V. and Sellwood, G. 2000, ApJ, 543, 704.
\bibitem[\protect\citeauthoryear{Ferguson et al.}{1998}]{Ferguson}
Ferguson, A., Wyse, R., Gallagher, J.  and Hunter, D. 1998, ApJ, 506, L19.
\bibitem[\protect\citeauthoryear{Kauffmann et al.}{2002}]{kauffmann}
Kauffmann, G. et al.  2002, astro-ph/0205070.
\bibitem[\protect\citeauthoryear{Klypin, Zhao and Somerville}{2002}]{Klypin}
Klypin, A., Zhao, H. and Somerville, R. 2002, ApJ, 573, 597.
\bibitem[\protect\citeauthoryear{Kranz, Slyz and Rix}{2002}]{Kranz}
Kranz, T., Slyz, A. and Rix, H. 2001, ApJ, 562, 164.
\bibitem[\protect\citeauthoryear{Lin and Pringle}{1987}]{Lin}
Lin, D. and Pringle, J. 1987, ApJ, 320, 87L.
\bibitem[\protect\citeauthoryear{Rocha-Pinto et al.}{2000}]{Rocha}
Rocha-Pinto, H., Scalo, J., Maciel, W. and Flynn, C. 2000, A\&A, 358, 869.
\bibitem[\protect\citeauthoryear{Silk}{2001}]{Silk}
Silk, J. 2001, MNRAS, 324, 313.
\bibitem[\protect\citeauthoryear{Silk and Norman}{1981}]{SilkN}
Silk, J. and Norman, C. 1981, ApJ, 247, 59.
\bibitem[\protect\citeauthoryear{Sommer-Larsen, Gotz and Portinari}{2002}]{sommer}
Sommer-Larsen, J., Gotz, M. and Portinari, L.  2002, astro-ph/0204366.
\bibitem[\protect\citeauthoryear{Strickland et al.}{2002}]{Strickland}
Strickland, D. et al. 2002, ApJ, 568, 689.
\bibitem[\protect\citeauthoryear{Thomas, Maraston and Bender}{2002}]{Thomas}
Thomas, D., Maraston, C. and Bender, R. 2002, astro-ph/0202166
\bibitem[\protect\citeauthoryear{van den Bosch}{2002}]{vandenBosch}
van den Bosch, F. 2002, MNRAS, 332, 456.
\bibitem[\protect\citeauthoryear{Vazdekis et al.}{2001}]{Vazdekis}
Vazdekis, A. et al. 2001, Ap.J., 551, L127.
\bibitem[\protect\citeauthoryear{Westera et al.}{2002}]{Westera}
Westera, P., Samlund, M., Buser, R. and Gerhard, O. 2002,
A\&A, submitted.
\bibitem[\protect\citeauthoryear{White \& Rees}{1978}]{White}
White, S. and Rees, M.   1978, MNRAS, 183, 341.
\end{thebibliography}
\end{document}